\documentclass[9pt,pra,twocolumn]{revtex4-2}
\usepackage{amsmath,amssymb,mathtools,braket,mathrsfs}

\begin{document}

\title{Simulating photon counting from dynamic quantum emitters \\ by exploiting zero-photon measurements}

\author{Stephen C. Wein}

\affiliation{Quandela, 7 Rue L\'{e}onard de Vinci, 91300 Massy, France}
\email{stephen.wein@quandela.com}

\begin{abstract}
Many applications of quantum optics demand delicate quantum properties of light carefully tailored to accomplish a specific task. To this end, numerical simulations of quantum light sources are vital for designing, characterizing, and optimizing quantum photonic technology. Here, I show that exploiting information hidden in zero-photon measurement outcomes provides an exponential speedup for time-integrated photon counting simulations, realizing eight orders of magnitude reduction in the time to compute six-photon detection probabilities while achieving ten orders of magnitude higher precision compared to the state of the art. This enables simulations of large photonic experiments with an unprecedented level of physical detail. It can accelerate the design of sources to generate photonic resource states for quantum sensing and measurement-based quantum computing while capturing realistic imperfections. It also establishes a general theoretical framework to study dynamic interactions between stationary qubits mediated by measurements of flying qubits, which can be used to model distributed quantum computing and quantum communication.
\end{abstract}

\maketitle

\section{Introduction}

Pulses of non-classical light serve as flying qubits for photonic quantum information processing \cite{flamini2018photonic}, are important resources for quantum sensing \cite{pirandola2018advances}, and are critical ingredients for a future quantum internet \cite{kimble2008quantum,simon2017towards,wehner2018quantum}. Combined with linear optics and photon-counting detectors, quantum states of light can be used to perform quantum computing \cite{knill2001scheme, kok2007linear, bartolucci2023fusion} and quantum communication \cite{o2009photonic}. They can be generated from classical pulses using non-linear processes such as parametric down-conversion \cite{couteau2018spontaneous} or emission from single quantum emitters \cite{aharonovich2016solid}. 

In particular for single emitters, the quantum dynamics of light-matter interaction play a large role in determining the quality of light produced \cite{kiraz2004quantum, gustin2020efficient, bracht2021swing}. Capturing these dynamics is also necessary to understand protocols that exploit matter degrees of freedom, such as the spin of a particle, to generate entangled states of light \cite{lindner2009proposal}, perform non-demolition measurements \cite{maffei2023energy}, or to serve as a quantum memory \cite{specht2011single}. However, achieving all of these applications to a sufficient quality for widespread commercialization requires a high level of engineering and optimization. It is thus crucial to develop accurate models and numerical simulation techniques that can provide critical feedback on current experiments and help develop proposals for next-generation quantum photonic devices.

Propagating pulses of light occupy the continuum of the electromagnetic field \cite{loudon2000quantum}. Hence, the physics of quantum photonic technology depends on continuous degrees of freedom, such as time or frequency, that are not fully resolved when measuring light. The standard way to simulate photon-counting measurements relies on computing field correlations in the time or frequency domain and subsequently integrating unresolved degrees of freedom to get the final measurement result. For example, simulating Hong-Ou-Mandel bunching \cite{hong1987measurement} of single-photon emission from a quantum dot requires integrating the arrival time of each photon at each detector to get the total coincidence probability \cite{kiraz2004quantum}. The consequence is that each photon-counting event contributes at least one dimension of integration, which scales poorly and quickly prohibits simulating the dynamics and measurement of more than a few interacting pulses of light. In some cases, multi-dimensional integrals can be analytically factored into lower-dimensional integrals \cite{grange2015cavity,wein2018feasibility}, which can alleviate the scaling problem. But, this still demands a fully time-resolved simulation and it must be hand-tailored to specific experimental setups.

To address this problem, I introduce a general method to simulate time-integrated quantities, such as photon-number probability distributions, without using multi-variable integration. The basic intuition is that it is relatively easy to simulate the probability of measuring zero photons using perturbation theory \cite{carmichael2009open, fischer2018particle, fischer2018scattering, wein2020analyzing}, because there is no arrival time and hence nothing to integrate. Counter-intuitively, this zero-photon probability can actually provide a lot of information about the state of light \cite{nunn2022modifying, nunn2023transforming}. Specifically, when using an inefficient detector, the zero-photon probability can be expressed as a linear combination of all photon-number probabilities weighted by powers of detector loss coefficients \cite{rossi2004photon,zambra2005experimental}. Thus, by indirectly simulating photon statistics by first computing zero-photon probabilities, it is possible to circumvent prohibitive multi-dimensional integration that hinders numerical simulations of many modern quantum optics experiments.

After making this connection, I show that the physical loss relation not only holds for zero-photon probabilities computed from source physics, but also for source conditional dynamics, multi-mode optical setups, and even mathematically extends to configurations of detectors with unphysical efficiencies. By focusing on zero-photon outcomes, a theoretical framework is uncovered that can considerably aid in the study and design of light produced by quantum sources. Notably, this framework leads to algorithms that provide an exponential speedup for photon-counting simulations and constitutes a robust numerical tool for studying a wide range of photonic experiments from boson sampling to spin-mediated cluster state generation using dynamic sources of light.

This paper is organized as follows. Section \ref{background} covers the background theory on the photon-number decomposition of an emitter quantum master equation. The results for single-mode scenario is given in Section \ref{singlemode} followed by the multi-mode extension in Section \Ref{multimode}. Section \ref{discussion} discusses applications and extensions of the method and Section \ref{conclusion} concludes the paper.

\section{Background}
\label{background}

Consider a quantum source of light that evolves following Markovian dynamics generated by a linear superoperator\footnote{I notate superoperators with a calligraphic font and operators using a hat. All superoperators act on everything to their right.} $\mathcal{L}$ called the Lindbladian \cite{manzano2020short}. The evolution of the density operator $\hat{\rho}$ is given by the Gorini–Kossakowski–Sudarshan–Lindblad master equation \cite{gorini1976completely, lindblad1976generators}
\begin{equation}
    \frac{d}{dt}\hat{\rho}(t) = \mathcal{L}(t)\hat{\rho}(t),
\end{equation}
for an initial state $\hat{\rho}(t_0)$. The solution is then given by $\hat{\rho}(t)=\mathcal{P}(t,t_0)\hat{\rho}(t_0)$, where the propagator is
\begin{equation}
\label{propagator}
    \mathcal{P}(t,t_0) = \mathscr{T}\exp\!\left[\int_{t_0}^t\mathcal{L}(t^\prime)dt^\prime\right],
\end{equation}
and where $\mathscr{T}$ orders time-dependent superoperators.

\begin{figure}
    \includegraphics[width=0.48\textwidth, trim=4 3 4 5, clip]{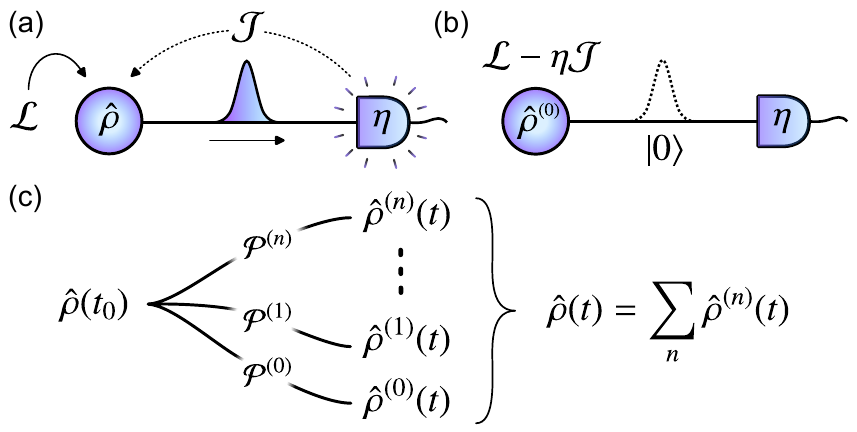}
    \caption{
\textbf{The photon-number decomposition.} (a) A source evolving with Markovian dynamics generated by the Lindbladian $\mathcal{L}$ emits a pulse of light collected into a single mode. The pulse is measured by an ideal photon-number resolving detector with efficiency $\eta$, which induces the linear superoperator $\mathcal{J}$ acting on the source density operator when a single photon is detected. (b) The absence of detected light conditions the source to evolve with dynamics governed by the zero-photon generator $\mathcal{L}^{(0)}_\eta=\mathcal{L}-\eta\mathcal{J}$. (c) The initial source density operator $\hat{\rho}(t_0)$ is decomposed into states $\hat{\rho}^{(n)}(t)=\mathcal{P}^{(n)}(t,t_0)\hat{\rho}(t_0)$ conditioned detecting $n$ photons between time $t_0$ and time $t$, where $p^{(n)}=\text{Tr}[\hat{\rho}^{(n)}]$ is the probability of detecting $n$ photons.}
    \label{figonemode}
\end{figure}

Suppose that the source emits a pulse of light that is monitored by a number-resolving detector with efficiency $\eta=1$ (see Fig. \ref{figonemode}a). The elementary problem is to simulate time-integrated quantities such as the probability $p^{(n)}$ of detecting $n$ photons during the detection window. Luckily for many source models, such as those satisfying a Heisenberg input-output relation \cite{gardiner1985input}, the detection of a photon at an instant $t$ implies that the source underwent an instantaneous state transition, described by a linear jump superoperator $\mathcal{J}(t)$, at the corresponding time of emission. This standard assumption \cite{breuer2002theory, kiraz2004quantum, wein2018feasibility, gustin2020efficient} allows the problem to be tackled using an intuitive open systems approach to quantum optics \cite{carmichael2009open} where the master equation can be decomposed into photon-number subspaces \cite{wein2020analyzing}.

The photon-number decomposition begins by constructing an effective master equation \cite{fischer2018particle} governed by a zero-photon generator (ZPG) 
\begin{equation}
    \mathcal{L}^{(0)}=\mathcal{L}-\mathcal{J}.
\end{equation}
In some cases, this ZPG can be rewritten as an effective non-Hermitian Hamiltonian, which is the primary object studied in quantum trajectories \cite{fischer2018scattering} and stochastic simulations \cite{breuer2002theory}.

The general solution $\mathcal{P}^{(0)}(t,t_0)$ to the effective master equation defined by a ZPG is similarly given by Eq.~(\ref{propagator}) when replacing $\mathcal{L}$ with $\mathcal{L}^{(0)}$. This propagator $\mathcal{P}^{(0)}$ describes the dynamics of the source conditioned on detecting zero photons (see Fig.~\ref{figonemode}b). By applying time-dependent perturbation theory, the full propagator $\mathcal{P}$ can be recovered using the mixed-state analog of the Dyson series
\begin{equation}
    \mathcal{P}=\sum_{n=0}^\infty \mathcal{P}^{(n)}
\end{equation}
to add back individual photon-counting events to the dynamics. The perturbations $\mathcal{P}^{(n)}(t,t_0)$ are source propagators conditioned on detecting $n$ photons between the initial time $t_0$ and the final time $t$, and they can be solved recursively \cite{wein2020analyzing} by 
\begin{equation}
    \mathcal{P}^{(n)}(t,t_0) = \int_{t_0}^t\mathcal{P}^{(0)}(t,t^\prime)\mathcal{J}(t^\prime)\mathcal{P}^{(n-1)}(t^\prime,t_0)dt^\prime.
\end{equation}
From this perspective, each photon-counting event adds a source jump $\mathcal{J}$ at some time $t^\prime$. The photon counting result is then given by integrating over all possible jump times $t^\prime$ between $t_0$ and $t$.

This photon-number decomposition (see Fig.~\ref{figonemode}c) provides the state of the source $\hat{\rho}^{(n)}(t)=\mathcal{P}^{(n)}(t,t_0)\hat{\rho}(t_0)$ given that $n$ photons have been detected between time $t_0$ and $t$, which occurs with the probability of $p^{(n)}(t)=\text{Tr}[\hat{\rho}^{(n)}(t)]$. However, the recursive solution, which can be evaluated using the scattering module in QuTiP \cite{johansson2012qutip}, implies that simulating $n$th-order time-integrated quantities requires solving an $n$-dimensional time integral. Hence, the time to compute $p^{(n)}$ in this way will scale exponentially, roughly following $\mathcal{O}\left({N_{dt}}^n\right)$ where $N_{dt}$ is the number of time steps needed to resolve the time dynamics.

\section{Results: Single-mode scenario}
\label{singlemode}

\begin{figure}
    \centering
    \includegraphics[width=0.49\textwidth, trim=12 0 0 5, clip]{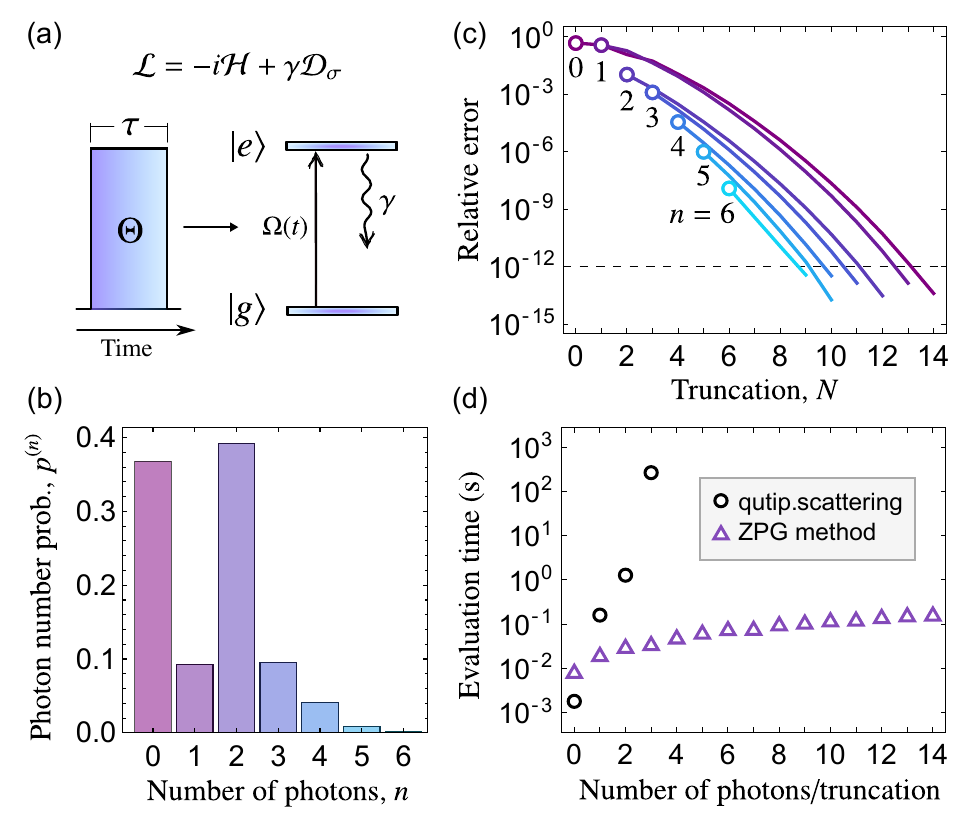}
    \caption{
\textbf{Scattering photons off a two-level emitter.} (a) A square pulse with temporal width $\tau$ and area $\Theta$ driving a two-level emitter with decay rate $\gamma$, whose evolution is governed by the Lindbladian $\mathcal{L}=-i\mathcal{H}+\gamma\mathcal{D}_\sigma$, where $\mathcal{H}\hat{\rho}=\Omega(t)[\hat{\sigma}+\hat{\sigma}^\dagger,\hat{\rho}]/2$, $\hat{\sigma}=\ket{g}\bra{e}$, and $\mathcal{D}_\sigma\hat{\rho}=\hat{\sigma}\hat{\rho}\hat{\sigma}^\dagger-\{\hat{\sigma}^\dagger\hat{\sigma},\hat{\rho}\}/2$. The zero-photon generator (ZPG) is then $\mathcal{L}^{(0)}_z=\mathcal{L}-\eta(z)\mathcal{J}$, where $\mathcal{J}\hat{\rho}=\gamma\hat{\sigma}\hat{\rho}\hat{\sigma}^\dagger$ and $\eta(z)=1-z^{-1}$. (b) Exact photon number probabilities $p^{(n)}$ for $\tau=2\gamma^{-1}$ and $\Theta=10\pi$ using analytic integration. (c) Convergence of the simulated distribution to the exact solution with increasing truncation $N$ of ZPG sampling points. (d) Numerical simulation time using the QuTiP scattering module (circles) for 2 significant digits of precision compared to the ZPG method (triangles) with up to 12 significant digits. }
    \label{fig2}
\end{figure}

To avoid the unfavorable scaling imposed by the recursive solution, one can exploit the relationship between the detector efficiency and the ZPG. If a photon is detected with probability $\eta$, the ZPG becomes \begin{equation}
    \mathcal{L}^{(0)}_\eta=\mathcal{L}-\eta\mathcal{J}.
\end{equation}
At the level of probabilities, the lossy zero-photon measurement outcome $p^{(0)}_\eta$ now not only depends on the true zero-photon probability $p^{(0)}$, but also on all higher-order probabilities $p^{(n)}$ multiplied by an appropriate detector loss coefficient. That is, one can infer that the loss relation must remain valid
\begin{equation}
    p^{(0)}_\eta = \sum_{n=0}^\infty (1-\eta)^np^{(n)}.
\end{equation}
Although this expression is a known probability generating function in linear optical optics \cite{rossi2004photon,zambra2005experimental, bulmer2022threshold}, we can now see that it is directly connected to the source dynamics through the solution to the ZPG. Notably, this realization is impossible if taking $\eta=1$ so that the ZPG can be reduced to an effective non-Hermitian Hamiltonian. 

Assuming the pulse has finite energy, there exists an $N$ such that $p^{(n)}\simeq 0$ for $n > N$. Then, by evaluating $p^{(0)}_\eta$ for $N$ unique values of $\eta$ along with $p^{(0)}_0=1$, all $p^{(n)}$ up to $n=N$ can be estimated by inverting the loss relation, as has been realized experimentally \cite{rossi2004photon,zambra2005experimental}. Most importantly, this indirect approach to obtain all non-negligible $p^{(n)}$ can be accomplished by solving the ZPG just $N$ times for different $\eta$. When neglecting the inversion step, which for reasonable $N$ is negligible compared to solving the dynamics, this leads to a linear scaling $\mathcal{O}\left(NN_{t}\right)$ and hence an exponential speedup. Here, $N_{t}$ is the number of time steps needed to solve the effective master equation until time $t$, and this can even be much smaller than $N_{dt}$ for time-independent evolution.

Arriving at the loss relation from physical arguments alone is not very satisfactory. In fact, as shown in Appendix~\ref{appendixA}, this loss relation outlined above is just a special case of a more general mathematical relation that extends to the conditional states $\hat{\rho}^{(n)}$ and propagators $\mathcal{P}^{(n)}$ of the source for any complex $\eta$, not just those bound to physical detector efficiencies $0\leq\eta\leq 1$. Hence, the first main result of this work is that the general solution 
\begin{equation}
\mathcal{G}_z(t,t_0)=\mathscr{T}\exp\!\left[\int_{t_0}^t\mathcal{L}^{(0)}_z(t^\prime)dt^\prime\right]
\end{equation}
to the effective master equation defined by a ZPG of the form 
\begin{equation}
    \mathcal{L}^{(0)}_z=\mathcal{L}-(1-z^{-1})\mathcal{J}
\end{equation}
is equal to the $\mathcal{Z}$-transform of the set of conditional propagators
\begin{equation}
\mathcal{G}_z=\mathcal{Z}\{\mathcal{P}^{(n)}\} \equiv \sum_{n=0}^\infty\mathcal{P}^{(n)}z^{-n}
\end{equation}
for $z\in\mathbb{C}$ and $z\neq 0$. Thus, the decomposition is obtained by the inverse transform $\mathcal{P}^{(n)}=\mathcal{Z}^{-1}\{\mathcal{G}_z\}$ for a set of unique $z$. In short, the proof involves taking the $n$th derivative of $\mathcal{G}_z$ with respect to $z^{-1}$ and then showing that $z\rightarrow \infty$ provides the lossless propagator $\mathcal{P}^{(n)}$. It follows from linearity that $\mathcal{G}_z\hat{\rho}(t_0)=\mathcal{Z}\{\hat{\rho}^{(n)}\}$ and $\text{Tr}[\mathcal{G}_z\hat{\rho}(t_0)]=\mathcal{Z}\{p^{(n)}\}$. The original loss relation is then recovered from this latter expression by setting $z=(1-\eta)^{-1}$.

A major advantage of the $\mathcal{Z}$-transform approach is that one need not restrict $\eta$ to be physical. This is because values of $\eta$ less than one lead to extremely small loss coefficients for large $n$, resulting in a loss of precision during the inversion step. Instead, the set of $z$ can now be chosen as roots of unity, $z^N=1$. Then, photon counting outcomes become encoded in the phase of unphysical zero-photon generating probabilities. In this case, the $\mathcal{Z}$-transform becomes a discrete Fourier transform and so $\mathcal{Z}^{-1}$ can be implemented using the numerically stable and optimized fast Fourier transform (FFT) algorithm \cite{seron2022efficient}.

The ability to quickly compute states and channels of the source conditioned on photon-counting outcomes is an additional major advantage of this approach that has a wide range of applications to study and design quantum devices. However, to demonstrate the exponential speedup I will focus on computing photon-counting probabilities.

Consider the textbook example of a two-level emitter driven by a square excitation pulse (see Fig. \ref{fig2}a), for which there is an analytic solution for $p^{(n)}$ \cite{fischer2018scattering}. To best illustrate the method, I choose an excitation pulse with an integrated area of $\Theta=10\pi$ and a temporal width $\tau=2\gamma^{-1}$ of twice the emitter lifetime $\gamma^{-1}$, so that the distribution is both non-classical and non-negligible up to $p^{(6)}$ (see Fig. \ref{fig2}b). The QuTiP package \cite{johansson2012qutip} can then be used to solve the ZPG and the photon number probabilities are reconstructed using an FFT. Figure~\ref{fig2}c shows that the simulated probability distribution converges exponentially to the analytically exact solution when increasing the truncation $N$, such that the relative error is $<10^{-12}$ for $p^{(n)}$ up to $n=6$ for $N=14$. In addition, Figure \ref{fig2}d shows an exponential speedup over the recursive integration method implemented using the QuTiP scattering module \cite{johansson2012qutip, fischer2018scattering}. Remarkably, extrapolating the recursive integration approach to $n=6$ indicates that an evaluation time of 150 days is needed to reach two significant digits of precision. On the other hand, using the ZPG along with an FFT accomplishes the task to a precision of 12 significant digits in just 100 ms.

The exponential speedup in the simulation of single-mode photon statistics enabled by the ZPG can be used to simulate figures of merit of quantum light sources and optimize control parameters, such as excitation pulse shape. However, to simulate contemporary large scale photonic experiments that may involve linear optical circuits and other optical elements, it is necessary to broaden the concept to include multiple sources producing emission into multiple possible modes of light.

\section{Results: Multi-mode scenario}
\label{multimode}

\begin{figure*}
    \includegraphics[width=\textwidth, trim=4 0 0 3, clip]{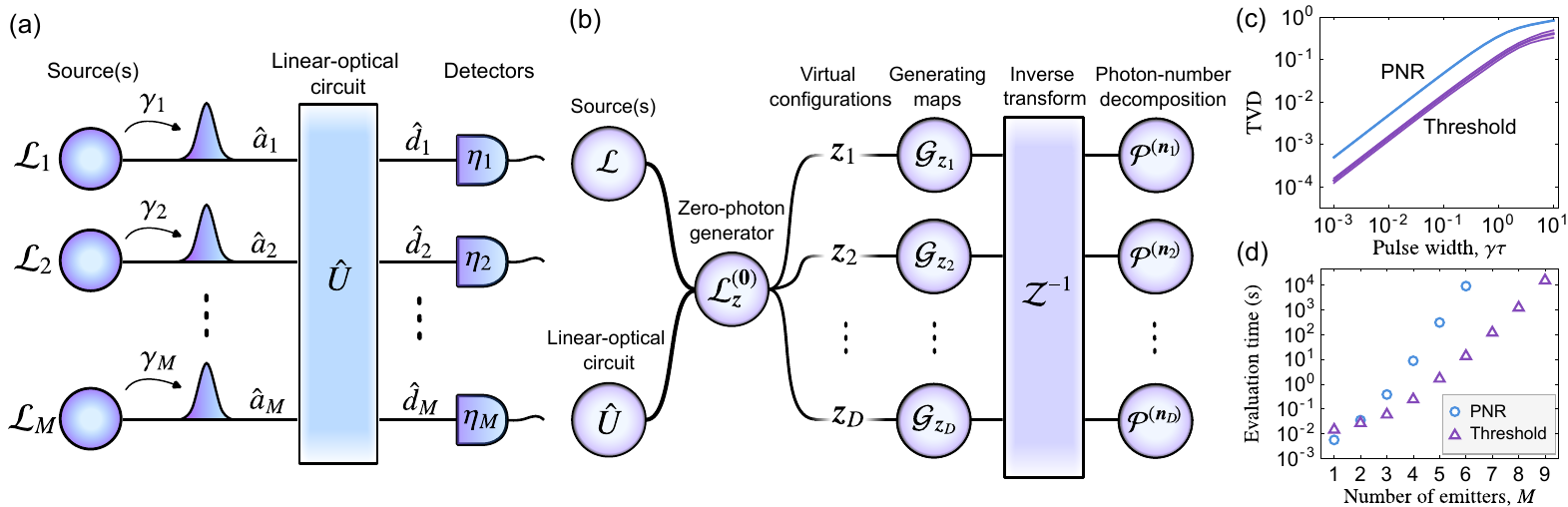}
    \caption{
\textbf{Multi-mode photon-number decomposition.} (a) An ensemble of sources, each described by a Lindbladian $\mathcal{L}_i$, emit pulses into their respective collection modes $\hat{a}_i$ at a rate $\gamma_i$. The pulses pass through a linear-optical circuit described by a unitary matrix $\hat{U}$. Each output $
    \hat{d}_i$ is monitored by a detector with efficiency $
    \eta_{i}$. (b) The absence of detection conditions the sources to evolve following the zero-photon generator (ZPG) $\smash{\mathcal{L}^{(\boldsymbol{0})}_{\boldsymbol{z}}}$. The ZPG is solved for $D$ unique virtual configurations $\boldsymbol{z}$, where $D$ is the number of outcomes $\boldsymbol{n}$ with non-negligible probability $p^{(\boldsymbol{n})}$. Applying the inverse transform $\mathcal{Z}^{-1}$ to the resulting set of generating maps $\{\mathcal{G}_{\boldsymbol{z}}\}$, states $\{\mathcal{G}_{\boldsymbol{z}}\hat{\rho}(t_0)\}$, or probabilities $\{\text{Tr}[\mathcal{G}_{\boldsymbol{z}}\hat{\rho}(t_0)]\}$ provides the set of conditional propagators $\{\mathcal{P}^{(\boldsymbol{n})}\}$, states $\{\hat{\rho}^{(\boldsymbol{n})}\}$, or probabilities $\{p^{(\boldsymbol{n})}\}$, respectively. (c) Average total variation distance (TVD) relative to perfect single-photon interference for emission from $M=4$ identical two-level emitters each driven by a square pulse with area $\Theta=\pi$ and a varying width $\tau$. The TVD for both photon-number resolved (PNR) and threshold detection distributions converge to zero as $\tau\rightarrow 0$. The curve thickness represents the standard deviation of the TVD over 10 Haar-random $\hat{U}$. (d) The time needed to simulate a full PNR or threshold detection distribution in Python as a function of the number of two-level emitters producing single photons.}
    \label{figmapping}
\end{figure*}

To generalize the method to one or more sources producing light collected into $M$ modes that are each monitored by a detector, we first construct the associated multi-mode ZPG \cite{wein2020analyzing}
\begin{equation}
\mathcal{L}_{\boldsymbol{z}}^{(\boldsymbol{0})} = \mathcal{L} - \boldsymbol{\eta}(\boldsymbol{z})\cdot\boldsymbol{\mathcal{J}} 
\end{equation}
where $\smash{\boldsymbol{\mathcal{J}}}=(\mathcal{J}_1,\dots,\mathcal{J}_M)$ is a vector of jump superoperators $\mathcal{J}_j$ describing the action on the source when detecting a photon by the $j$th detector, and $\boldsymbol{\eta}=(\eta_1,\dots,\eta_M)$ is a vector of corresponding virtual detector efficiencies $\eta_j = 1-z_j^{-1}$ for $z_j\in\mathbb{C}$. Since adding additional detectors only adds linear perturbations to the ZPG, the $\mathcal{Z}$-transform extends in a transparent way leading to the second main result of this work
\begin{equation}
\label{ZPG}  \mathscr{T}\exp\left[\int_{t_0}^t\mathcal{L}^{(\boldsymbol{0})}_{\boldsymbol{z}}(t^\prime)dt^\prime\right]= \sum_{\boldsymbol{n}}\mathcal{P}^{(\boldsymbol{n})}(t,t_0)\prod_{j=1}^Mz_j^{-n_j},
\end{equation}
where $\boldsymbol{n}=(n_1,\dots,n_M)$ is the vector of detected photon numbers and $\mathcal{P}^{(\boldsymbol{n})}$ is the propagator conditioned on observing $\boldsymbol{n}$. In analogy with the loss relation, the term $\prod_{j=1}^Mz_j^{-n_j}$ represents the conditional probability of detecting zero photons given the photon pattern $\boldsymbol{n}$.

To elaborate on a specific scenario, consider a system of $M$ independent classically-driven quantum sources (see Fig. \ref{figmapping}a). Each source is governed by a Lindbladian $\mathcal{L}_i$ and satisfies a Heisenberg input-output relation $\hat{a}_i=\sqrt{\gamma_i}\hat{c}_i+\hat{a}_{i,\mathrm{in}}$ arising from a linear dipole interaction in the Markovian limit \cite{gardiner1985input}, where $\hat{a}_i$ is the mode collecting emission, and $\hat{c}_i$ is the system operator coupled to the electromagnetic continuum with rate $\gamma_i$. The operator $\hat{a}_{i,\mathrm{in}}$ describes the quantum fluctuations of the electromagnetic vacuum input to the $i$th source, which is inconsequential when simulating photon-counting measurements \cite{carmichael2009open}. Also, as is typical of boson-sampling type experiments, suppose there is a linear-optical unitary transformation $\hat{U}$ on the collection modes producing output modes $\hat{d}_j = \sum_{i}{U}_{\!ji}\hat{a}_i$ that are each monitored by a detector. 

By choosing to decompose the dynamics using jump superoperators $\mathcal{J}_{j}$ that describe the detection of a photon at the $j$th detector \emph{after} the unitary transformation, the multi-mode ZPG can be rewritten as
\begin{equation}
\label{zerophotoneq}
    \mathcal{L}_{\boldsymbol{z}}^{(\boldsymbol{0})} = \mathcal{L} - \boldsymbol{\mathcal{J}}^+\cdot\hat{\eta}^\prime(\boldsymbol{z})\cdot\boldsymbol{\mathcal{J}}^-,
\end{equation}
where $\mathcal{L}=\sum_{i=1}^M\mathcal{L}_i$, and $\boldsymbol{\mathcal{J}}^\pm=(\mathcal{J}_1^\pm,\dots,\mathcal{J}_M^\pm)$ with $\mathcal{J}^-_i\hat{\rho}=\sqrt{\gamma_i}\hat{c}_i\hat{\rho}$ and $\mathcal{J}^+_i\hat{\rho}=\sqrt{\gamma_i}\hat{\rho}\hat{c}_i^\dagger$. The matrix $\hat{\eta}^\prime(\boldsymbol{z})=\hat{U}^\dagger\hat{\eta}(\boldsymbol{z})\hat{U}$ is the unitary transformation of the diagonal matrix $\hat{\eta}(\boldsymbol{z})$ of virtual efficiencies $\boldsymbol{\eta}(\boldsymbol{z})$. Note that a source $k$ can also produce uncorrelated vacuum by setting $\gamma_k=0$ and neglecting $\mathcal{L}_k$.

Interestingly, from the perspective of the source, the unitary transformation acts on the detector efficiencies rather than the modes of light. In addition, this multi-mode ZPG strongly resembles a Hamiltonian of a coupled many-body system with potential long-range two-body interactions. By expanding the coupling term, it is apparent that each source experiences a local zero-photon shift $\mathcal{L}_i-{\eta}^\prime_{ii}\mathcal{J}^+_i\mathcal{J}^-_i$ and there is a conditional coupling ${\eta}_{ij}^\prime\mathcal{J}^+_i\mathcal{J}_j^- \!+ {\eta}_{ji}^\prime\mathcal{J}^-_i\mathcal{J}_j^+$ that depends critically on the efficiency matrix $\hat{\eta}$ and the unitary transformation $\hat{U}$. For example, if $\hat{\eta}$ is the identity, then the observation of no photons implies each input was vacuum, and hence all sources must each follow their local zero-photon evolution governed by $\smash{\mathcal{L}^{(0)}_i}$. If $\hat{\eta}$ is zero, then the zero-photon measurement provides no information and each source independently evolves following $\mathcal{L}_i$. Otherwise, provided that $\hat{\eta}$ and $\hat{U}$ do not commute, zero-photon measurement outcomes generate correlations in the source dynamics.

Consider again the case of a two-level emitter driven by a square excitation pulse, but now with integrated area $\Theta=\pi$ so that its emission converges to an ideal single photon as the excitation pulse width tends to zero $\tau\rightarrow 0$. To demonstrate the ZPG method for multi-mode simulations (see Fig.~\ref{figmapping}b), I evaluate zero-photon probabilities of $M$ such emitters and use a multi-dimensional FFT to reconstruct the photon detection statistics following Eq.~(\ref{ZPG}). For various pulse widths $\tau$ and unitary transformations $\hat{U}$, I compute the total variation distance (TVD) of the simulated probability distribution relative to the exact distribution computed using Perceval \cite{heurtel2023perceval} (see code availability). Figure~\ref{figmapping}c shows that the TVD averaged over 10 Haar random $4\times 4$ unitary matrices approaches zero as the pulse width decreases, verifying that the method reproduces exact single-photon quantum interference patterns in the limit that the dynamics of each source leads to the emission of ideal single photons. Thus, the multi-mode ZPG correctly captures correlations due to quantum optical interference.

The time to simulate the full distribution for $M$ two-level emitters and $M$ detectors increases exponentially (see~Fig.~\ref{figmapping}d), as expected due to the exponentially increasing Hilbert space size and number of outcomes. However, the speedup provided by the ZPG method enables, for the first time, the simulation of exact time-integrated quantum dynamics, interference, and full photon-number resolved probability distribution of pulsed emission from up to six emitters in less than three hours on a laptop using Python. Preliminary work also suggests that optimization using Julia or C++ could decrease this time by up to two orders of magnitude.

\section{Discussion}
\label{discussion}

The ZPG method has multiple extensions and applications. Notably, the ZPG can be evaluated independently for each configuration $\boldsymbol{z}$, allowing for embarrassingly parallel computation and potential implementations using high performance computing. The set of $\boldsymbol{z}$ and corresponding $\mathcal{Z}$-transform can also be designed to efficiently provide other quantities that can be written as a function of $p^{(\boldsymbol{n})}$. For example, as shown in Appendix~\ref{appendixB}, the method can provide threshold detection probabilities directly \cite{bulmer2022threshold}, which drastically reduces simulation times (see~Fig.~\ref{figmapping}d) and is relevant for state-of-the-art photonic devices \cite{maring2023general}. 

Exploiting threshold detection along with a ZPG can also be used to derive efficient algorithms to directly simulate figures of merit for single-photon sources such as brightness, single-photon purity, and indistinguishability without integrating multi-time field correlation functions. In addition, the method is not limited to photon counting simulations. By including a local oscillator explicitly as a source \cite{carmichael2009open}, the method can be used to simulate homodyne measurements of time-integrated Wigner functions at individual points in phase space \cite{banaszek1999direct} by setting $z=-1$ so that the $\mathcal{Z}$-transform becomes the parity summation.

The method is fully compatible with the SLH framework for quantum cascaded networks \cite{combes2017slh}, which enables the simulation of sources with non-vacuum input fields \cite{kiilerich2019input} or circuits containing non-linear materials. The unitary property of $\hat{U}$ can also be relaxed,  at no disadvantage, to take into account non-uniform circuit losses, which are notoriously neglected in noisy boson sampling simulations. 

Simulated measurements can involve many different degrees of freedom of light emitted by one or more sources, such as polarization, frequency, spatial mode, and time bin. Degrees of freedom can be binned together to accurately represent experimental setups while drastically reducing simulation time \cite{seron2022efficient}. Moreover, since the method gives the dynamics conditioned on photon-counting measurements, it can be used to simulate realistic conditional quantum channels for spin-photon entanglement \cite{coste2022high} and photon-heralded spin-spin entanglement protocols for quantum communication \cite{atature2018material, wein2020analyzing,pompili2021realization}.

It is worth noting that the ZPG and corresponding $\mathcal{Z}$-transform can also be used to simulate optical systems without the need to explicitly model the source of light. In this case, the ZPG can be treated as a purely phenomenological object that captures the properties and evolution of light, such as decoherence. Since, for a fixed ZPG dimension and mode number, the approach enables photon-counting simulations that scale linearly with the number of photons, this framework can be used to simulate single-mode pulses of light composed of potentially thousands of photons while still capturing quantum properties. As such, it may serve as an attractive framework to explore quantum-to-classical transitions in optics.

\section{Conclusion}
\label{conclusion}

By exploiting source physics conditioned on zero-photon measurement outcomes, it is possible to circumvent multi-dimensional integration when simulating time-integrated photon counting. This provides an exponential computational speedup for simulating photon-counting experiments using time-dynamic quantum systems, which has a broad range of applications in quantum photonics. Further studies could extend the concept of a ZPG to include non-Markovian dynamics, linear-optical circuits that include delay lines, and measurement feed-forward.

The ZPG defines an equation of motion that can simulate quantum information processing using stationary qubits, flying qubits, or both in a hybrid approach. Thus it connects two physically very different quantum technology paradigms. It is promising to develop an analogy between the ZPG and the Hamiltonian dynamics of many-qubit systems to uncover algorithms that exploit noise to solve the ZPG more efficiently, such as tensor network techniques \cite{orus2019tensor,oh2021classical}. In addition, the coupling between sources during the photonic measurement, and hence the correlations built up over time, depends on the efficiency, unitary transformation, and coherence between emission from each source. Therefore the ZPG also has features in line with classical simulability of boson sampling problems \cite{moylett2019classically,garcia2019simulating,mezher2022assessing}, which provides a new theoretical perspective that could lead to further studies on the complexity and quantum advantage for photonic quantum information processing.

\section*{Acknowledgements}
This work was supported by the European Innovation Council (EIC) Accelerator program through the Scalable Entangled-Photon based Optical Quantum Computers (SEPOQC) grant, project number 190188855. I would like to thank Paul Hilaire and Neil Sinclair for reading the manuscript and providing feedback; Shane Mansfield, Rawad Mezher, and Emilio Annoni for discussions about theory; Fabien Thollot for help with clarification of concepts; Sharon David and Albert Adiyatullin for testing implementations; and Jean Senellart, Nicolas Heurtel, Timoth\'{e}e Goubault de Brugi\`{e}re, Raksha Singla, Valentin Guichard, Hubert Lam, Nadia Belabas, Pascale Senellart, and H\'{e}l\`{e}ne Ollivier for thought-provoking questions and comments.

\section*{Code Availability}
All code used to produce the numerical results will be made available in a public repository upon publication.

\section*{Disclosures}
SCW is an employee of Quandela, who has submitted a patent concerning the implementation of some methods introduced in this paper.

% \section*{Supplementary Material}
% Three notebooks are provided to replicate all the presented numerical results. Code~1 is a Mathematica notebook that evaluates the exact analytic expression for scattering probabilities. Code~2 is a Python notebook that implements the ZPG method to simulate single-mode and multi-mode probability distributions. Code~3 is a Julia notebook that computes threshold detection statistics using a specialized noise model and an optimized evaluation approach.

\appendix
\section{Photon-number decomposition using a Z-transform}
\label{appendixA}

To demonstrate that the zero-photon conditional propagator $\mathcal{P}^{(0)}_\eta(t,t_0)=\mathscr{T}\exp[\int_{t_0}^t\mathcal{L}_\eta^{(0)}(t^\prime)dt^\prime]$ defined by the ZPG $\mathcal{L}_\eta^{(0)}(t)=\mathcal{L}(t)-\eta\mathcal{J}(t)$ is equal to the generating map $\mathcal{G}_z=\mathcal{Z}\{\mathcal{P}^{(n)}\}=\sum_{n=0}^\infty\mathcal{P}^{(n)}z^{-n}$ for $z=(1-\eta)^{-1}$, we can equate each coefficient of the polynomial by showing that $d^n\mathcal{P}^{(0)}_\eta/dL^n|_{L\rightarrow0} = n!\mathcal{P}^{(n)}$ where $L=1-\eta$ is a complex loss coefficient. Note we already have $\mathcal{P}^{(n)}_\eta|_{L\rightarrow0}=\mathcal{P}^{(n)}$ for all $n$ by definition and so it suffices to show that $d\mathcal{P}_\eta^{(n)}/dL=(n+1)\mathcal{P}^{(n+1)}_{\eta}$ for all $n$. 

To proceed we can first consider the case where $\mathcal{L}$ and $\mathcal{J}$ do not depend on time. Then $\mathcal{P}_\eta^{(0)}(t, t_0)=\exp[(t-t_0)(\mathcal{L}-\eta\mathcal{J})]$. For the base case showing $n=0$ implies $n=1$, we can make use of the Wilcox formula for the exponential map:
\begin{equation}
    \frac{d}{dx}e^{A(x)} = \int_{0}^1e^{\alpha A(x)}\frac{dA(x)}{dx}e^{(1-\alpha)A(x)}d\alpha
\end{equation}
to find that $d\mathcal{P}^{(0)}_\eta(t,t_0)/dL$ is
\begin{equation}
\begin{aligned}
        \!(t-t_0)\!\!\int_{0}^1\!\!e^{\alpha (t-t_0)(\mathcal{L}-\eta\mathcal{J})}\mathcal{J}e^{(1-\alpha)(t-t_0)(\mathcal{L}-\eta\mathcal{J})}d\alpha
\end{aligned}
\end{equation}
By substituting $\alpha(t-t_0) = t-t^\prime$ we get \begin{equation}
\begin{aligned}
        \frac{d}{dL}\mathcal{P}^{(0)}_{\eta}(t,t_0) &= \int_{t_0}^{t}e^{(t-t^\prime)(\mathcal{L}-\eta\mathcal{J})}\mathcal{J}e^{(t^\prime-t_0)(\mathcal{L}-\eta\mathcal{J})}dt^\prime\\
        &= \int_{t_0}^{t}\mathcal{P}^{(0)}_\eta(t,t^\prime)\mathcal{J}\mathcal{P}^{(0)}_\eta(t^\prime,t_0)dt^\prime\\
        &=\mathcal{P}_\eta^{(1)}(t,t_0).
\end{aligned}
\end{equation}
Now, if we assume $d\mathcal{P}^{(n-1)}_\eta/dL = n\mathcal{P}_\eta^{(n)}$, then
\begin{widetext}
\begin{equation}
\begin{aligned}
    \frac{d}{dL}\mathcal{P}^{(n)}_\eta(t,t_0) &= \frac{d}{dL}\int_{t_0}^t\mathcal{P}^{(0)}_\eta(t,t^\prime)\mathcal{J}\mathcal{P}^{(n-1)}_\eta(t^\prime,t_0)dt^\prime\\
    &=\int_{t_0}^t\frac{d\mathcal{P}^{(0)}_\eta(t,t^\prime)}{dL}\mathcal{J}\mathcal{P}^{(n-1)}_\eta(t^\prime,t_0)dt^\prime + \int_{t_0}^t\mathcal{P}^{(0)}_\eta(t,t^\prime)\mathcal{J}\frac{d\mathcal{P}^{(n-1)}_\eta(t^\prime,t_0)}{dL}dt^\prime\\
    &=\int_{t_0}^t\mathcal{P}^{(1)}_\eta(t,t^\prime)\mathcal{J}\mathcal{P}^{(n-1)}_\eta(t^\prime,t_0)dt^\prime + n\int_{t_0}^t\mathcal{P}^{(0)}_\eta(t,t^\prime)\mathcal{J}\mathcal{P}^{(n)}_\eta(t^\prime,t_0)dt^\prime\\
    &=(n+1)\mathcal{P}^{(n+1)}_\eta(t,t_0).
\end{aligned}
\end{equation}
\end{widetext}
The last step combining the two terms makes use of the relation
\begin{equation}
\int_{t_0}^t\mathcal{P}^{(n)}(t,t^\prime)\mathcal{J}\mathcal{P}^{(k)}(t^\prime,t_0)dt^\prime =\mathcal{P}^{(n+k+1)}(t,t_0),
\end{equation}
a proof of which is in the appendix of Ref. \cite{wein2021modelling}.

To extend this to the time-dependent case, we can divide the total time interval into $N$ piece-wise time-independent parts each of length $dt=(t-t_0)/N$, beginning at time $t_{i-1}$ and ending at time $t_i$. Since each $\mathcal{G}_z(t_{i},t_{i-1})$ satisfies an effective master equation, we simply have $\mathcal{G}_z(t,t_0)=\prod_{i=1}^N\mathcal{G}_z(t_{i},t_{i-1})$. Then, we can substitute the time-independent solution and regroup terms based on the total number of photons
\begin{equation}
\begin{aligned}
\mathcal{G}_z(t,t_0)&= \prod_{i=1}^N\sum_{n=0}^\infty\mathcal{P}^{(n)}(t_i,t_{i-1})z^{-n}
\end{aligned}
\end{equation}
and so
\begin{equation}
\begin{aligned}
\prod_{i=1}^N\sum_{n=0}^\infty&\mathcal{P}^{(n)}(t_i,t_{i-1})z^{-n}=\mathcal{P}^{(0)}(t,t_0)\\
+& z^{-1}\sum_{i=1}^N\mathcal{P}^{(0)}(t,t_i)\mathcal{P}^{(1)}(t_i,t_{i-1})\mathcal{P}^{(0)}(t_{i-1},t_0)\\
+&\cdots,
\end{aligned}
\end{equation}
where $t_N=t$. Taking the limit $dt\rightarrow 0$, we can find that $\mathcal{P}^{(n)}(t_i,t_{i-1})$ for $n\geq 2$ are negligible compared to all combinations of $n$ single-photon propagators $\mathcal{P}^{(1)}$ among bins of vacuum $\mathcal{P}^{(0)}$. In addition, $\mathcal{J}$ becomes localized at $t_{i-1}\leq t^\prime\leq t_i$. So, substituting the definition of $\mathcal{P}^{(1)}$ and moving to the continuum limit we get
\begin{widetext}
\begin{equation}
\begin{aligned}
    \mathcal{G}_z(t,t_0)&=\mathcal{P}^{(0)}(t,t_0) + z^{-1}\sum_{i=1}^N\int_{t_{i-1}}^{t_{i}}\mathcal{P}^{(0)}(t,t_i)\mathcal{P}^{(0)}(t_i,t^\prime)\mathcal{J}(t_i)\mathcal{P}^{(0)}(t^\prime,t_{i-1})\mathcal{P}^{(0)}(t_{i-1},t_0)dt^\prime+\cdots\\
    &=\mathcal{P}^{(0)}(t,t_0) + z^{-1}\sum_{i=1}^N\int_{t_{i-1}}^{t_{i}}\mathcal{P}^{(0)}(t,t^\prime)\mathcal{J}(t_i)\mathcal{P}^{(0)}(t^\prime,t_0)dt^\prime+\cdots\\
    &=\mathcal{P}^{(0)}(t,t_0) + z^{-1}\int_{t_0}^{t}\mathcal{P}^{(0)}(t,t^\prime)\mathcal{J}(t^\prime)\mathcal{P}^{(0)}(t^\prime,t_0)dt^\prime+\cdots\\
    &=\sum_{n=0}^\infty \mathcal{P}^{(n)}(t,t_0)z^{-n}.
\end{aligned}
\end{equation}
\end{widetext}
Although I only illustrated the regrouping for the $n=1$ terms, the same argument applies to the regrouping of the $n\geq 2$ terms.

The decomposition has a straightforward extension to the multi-mode scenario \cite{wein2020analyzing}. The perturbative series becomes $\mathcal{P}=\sum_{\boldsymbol{n}}\mathcal{P}^{(\boldsymbol{n})}$ where
\begin{equation}
    \mathcal{P}^{(\boldsymbol{n}+\boldsymbol{e}_i)}(t,t_0) = \int_{t_0}^t\mathcal{P}^{(\boldsymbol{0})}(t,t^\prime)\mathcal{J}_i(t^\prime)\mathcal{P}^{(\boldsymbol{n})}(t^\prime,t_0)dt^\prime,
\end{equation}
where $\boldsymbol{e}_i$ is the $i$th unit vector. The zero-photon propagator $\mathcal{P}^{(\boldsymbol{0})}(t,t_0)$ is the solution to the effective master equation $d\hat{\rho}^{(\boldsymbol{0})}(t)/dt=\mathcal{L}^{(\boldsymbol{0})}(t)\hat{\rho}^{(\boldsymbol{0})}(t)$ where the ZPG is
\begin{equation}
   \mathcal{L}^{(\boldsymbol{0})}(t) = \mathcal{L}(t) - \sum_i\mathcal{J}_i(t).
\end{equation}
Multiplying a detector efficiency $\eta_i$ to each $\mathcal{J}_i$, the ZPG takes the form given in the main text. Since adding additional detectors only adds independent perturbations linearly to the ZPG, the proof of the single-mode scenario immediately extends due to the linearity of the derivative in the Wilcox formula.

\section{Threshold detection decomposition}
\label{appendixB}

Often measurements are performed where a detector `clicks' if it receives one or more photons. We denote probability of that the detector clicks as the brightness $\beta=\sum_{n=1}^\infty p^{(n)}=1-p^{(0)}$, where $p^{(0)}$ is the probability that the detector does not click. The conditional state associated with the threshold detection probability is then similarly given by the complement of the zero-photon conditional state: the bright conditional state $\hat{\beta}=\hat{\rho}-\hat{\rho}^{(0)}$. Even more generally, the associated bright propagation superoperator is $\mathcal{B}=\mathcal{P}-\mathcal{P}^{(0)}$ \cite{coste2022high}. In summary, we have $    \beta(t)=\text{Tr}[\hat{\beta}(t)]=\text{Tr}[\mathcal{B}(t,t_0)\hat{\rho}(t_0)]$ for initial state $\hat{\rho}(t_0)$ of the system. 

When there are multiple detectors, the threshold detection probabilities are more conveniently notated by $\beta^{(\boldsymbol{m})}$. Here, $\boldsymbol{m}$ is a vector of binary numbers where $1$ represents a threshold detection as opposed to $\boldsymbol{n}$ in $p^{(\boldsymbol{n})}$, which represents the vector of detected photon numbers. It is important to note that, unlike the single-mode case, the threshold detection probability distribution $\beta^{(\boldsymbol{m})}$ cannot be computed by $1-p^{(\boldsymbol{0})}$.

To recover the associated bright propagation superoperators $\mathcal{B}^{(\boldsymbol{m})}$ conditioned on the threshold detection outcome $\boldsymbol{m}$ from the ZPG, we can notice that there is a special case of the transform where each $z_i$ either tends to infinity (efficient limit, $L\rightarrow 0$) or tends to 1 (lossy limit, $L\rightarrow 1$). Then, $\prod_i z_i^{-n_i} \rightarrow \prod_{i} L_i^{n_i}$, where $L_i^{n_i}$ is either 1 or 0 (and $L_i^0\rightarrow 1$ for $L_i\rightarrow 0$). We can then see that $L_i^{n_i}=L_i$ if $n_i\geq 1$ and $L_i^{n_i}$=1 if $n_i=0$. Hence, all terms $\mathcal{P}^{(\boldsymbol{n})}\prod_{i} L_i^{n_i}$ that differ by some $n_i\neq 0$ will be identical and sum to the associated $\mathcal{B}^{(\boldsymbol{n})}\prod_{i} L_i^{n_i}$. In this particular case, the transformation can be inverted \cite{bulmer2022threshold} to obtain a solution for the threshold detection decomposition
\begin{equation}
\label{thresholdbeta}
    \mathcal{B}^{(\boldsymbol{m})} = \sum_{\boldsymbol{z}}\mathcal{G}_{\boldsymbol{z}}\prod_{i}(-1)^{m_i+L_i}(1-L_i)^{1-m_i},
\end{equation}
where $z_i=L_i^{-1}$ and $L_i=1-\eta_i$.

\end{document}